\documentclass[12pt]{article}
\usepackage{graphicx}
\usepackage{lscape}
\usepackage{citesort}
\usepackage{amssymb}
\usepackage{appendix}
\usepackage{multirow}

\begin{document}
\def\qq{\langle \bar q q \rangle}
\def\uu{\langle \bar u u \rangle}
\def\dd{\langle \bar d d \rangle}
\def\sp{\langle \bar s s \rangle}
\def\GG{\langle g_s^2 G^2 \rangle}
\def\Tr{\mbox{Tr}}
\def\figt#1#2#3{
        \begin{figure}
        $\left. \right.$
        \vspace*{-2cm}
        \begin{center}
        \includegraphics[width=10cm]{#1}
        \end{center}
        \vspace*{-0.2cm}
        \caption{#3}
        \label{#2}
        \end{figure}
	}
	
\def\figb#1#2#3{
        \begin{figure}
        $\left. \right.$
        \vspace*{-1cm}
        \begin{center}
        \includegraphics[width=10cm]{#1}
        \end{center}
        \vspace*{-0.2cm}
        \caption{#3}
        \label{#2}
        \end{figure}
                }

\def\ds{\displaystyle}
\def\beq{\begin{equation}}
\def\eeq{\end{equation}}
\def\bea{\begin{eqnarray}}
\def\eea{\end{eqnarray}}
\def\beeq{\begin{eqnarray}}
\def\eeeq{\end{eqnarray}}
\def\ve{\vert}
\def\vel{\left|}
\def\ver{\right|}
\def\nnb{\nonumber}
\def\ga{\left(}
\def\dr{\right)}
\def\aga{\left\{}
\def\adr{\right\}}
\def\lla{\left<}
\def\rra{\right>}
\def\rar{\rightarrow}
\def\lrar{\leftrightarrow}  
\def\nnb{\nonumber}
\def\la{\langle}
\def\ra{\rangle}
\def\ba{\begin{array}}
\def\ea{\end{array}}
\def\tr{\mbox{Tr}}
\def\ssp{{\Sigma^{*+}}}
\def\sso{{\Sigma^{*0}}}
\def\ssm{{\Sigma^{*-}}}
\def\xis0{{\Xi^{*0}}}
\def\xism{{\Xi^{*-}}}
\def\qs{\la \bar s s \ra}
\def\qu{\la \bar u u \ra}
\def\qd{\la \bar d d \ra}
\def\qq{\la \bar q q \ra}
\def\gGgG{\la g^2 G^2 \ra}
\def\q{\gamma_5 \not\!q}
\def\x{\gamma_5 \not\!x}
\def\g5{\gamma_5}
\def\sb{S_Q^{cf}}
\def\sd{S_d^{be}}
\def\su{S_u^{ad}}
\def\sbp{{S}_Q^{'cf}}
\def\sdp{{S}_d^{'be}}
\def\sup{{S}_u^{'ad}}
\def\ssp{{S}_s^{'??}}

\def\sig{\sigma_{\mu \nu} \gamma_5 p^\mu q^\nu}
\def\fo{f_0(\frac{s_0}{M^2})}
\def\ffi{f_1(\frac{s_0}{M^2})}
\def\fii{f_2(\frac{s_0}{M^2})}
\def\O{{\cal O}}
\def\sl{{\Sigma^0 \Lambda}}
\def\es{\!\!\! &=& \!\!\!}
\def\ap{\!\!\! &\approx& \!\!\!}
\def\ar{&+& \!\!\!}
\def\ek{&-& \!\!\!}
\def\kek{\!\!\!&-& \!\!\!}
\def\cp{&\times& \!\!\!}
\def\se{\!\!\! &\simeq& \!\!\!}
\def\eqv{&\equiv& \!\!\!}
\def\kpm{&\pm& \!\!\!}
\def\kmp{&\mp& \!\!\!}
\def\mcdot{\!\cdot\!}
\def\erar{&\rightarrow&}


\def\simlt{\stackrel{<}{{}_\sim}}
\def\simgt{\stackrel{>}{{}_\sim}}

\def\olra{\stackrel{\leftrightarrow}}
\def\ola{\stackrel{\leftarrow}}
\def\ora{\stackrel{\rightarrow}}


\title{
         {\Large
                 {\bf
Magnetic dipole moments of the negative parity $J^{PC}=2^{--}$ mesons in QCD
                 }
         }
      }

\author{\vspace{1cm}\\
{\small T. M. Aliev$^1$ \thanks {e-mail: taliev@metu.edu.tr}~\footnote{permanent address:Institute of
Physics,Baku,Azerbaijan}\,\,,
T. Barakat$^2$ \thanks {e-mail:
tbarakat@KSU.EDU.SA}\,\,,
M. Savc{\i}$^1$ \thanks
{e-mail: savci@metu.edu.tr}} \\
{\small $^1$\,Physics Department, Middle East Technical University,
06800 Ankara, Turkey }\\
{\small $^2$\,Physics and Astronomy Department, King Saud University, Saudi
Arabia}}

\date{}

\begin{titlepage}
\maketitle
\thispagestyle{empty}

\begin{abstract}
Magnetic dipole moments of the negative parity light and heavy tensor
mesons are calculated within the light cone QCD sum rules method. The
results are compared with the positive parity counterparts of the
corresponding tensor mesons. The results of the analysis show that
the magnetic dipole moments of the negative parity light mesons are
smaller compared to those of the positive parity mesons. Contrary to the
light meson case, magnetic dipole moments of the negative parity heavy
mesons are larger than the ones for the positive parity mesons.  
\end{abstract}

\vspace{1cm}
~~~PACS number(s): 11.55.Hx, 13.25.Jx, 13.40.Em
\end{titlepage}

\section{Introduction}

The study of the spectroscopy of particles play critical role for
understanding the dynamics of quantum chromodynamics (QCD), both at large
and short distances. According to the conventional quark model, the
particles are characterized by the $J^{PC}$ quantum numbers $P=(-1)^{L+1}$
and $C=(-1)^{L+S}$, where $L$ and $S$ are the orbital angular momentum
and the total spin, respectively. The spectroscopy of the particles with the
quantum numbers $J^{PC}=0^{\pm +}$, $1^{\pm -}$, $1^{++}$ are widely
investigated elsewhere in the literature. The mass and residues of light
tensor mesons are studied firstly in \cite{Rugr01} in the framework of the
QCD sum rules method. Later similar studies are extended for the strange
tensor mesons in \cite{Rugr02}. The masses and decay constants of the ground
states of the heavy $\chi_{Q_2}$ tensor mesons are investigated within the
same framework in \cite{Rugr03}. The relevant quantities in understanding
the internal structure of the mesons and baryons are their electromagnetic
form factors and multipole moments, such as the dipole moments. The dipole
moments for the heavy and light tensor mesons with the quantum number
$J^P=2^+$ are investigated in the framework of the light cone QCD sum rules
in \cite{Rugr04} and \cite{Rugr05}, respectively. The negative parity tensor
mesons have received less attention. The first attempt has recently been
made to calculate the mass and decay constants of the negative parity
$\bar{q}q$, $\bar{q}s$, $\bar{s}s$, $\bar{q}c$, $\bar{s}c$, $\bar{q}b$,
$\bar{s}b$, and $\bar{c}b$ tensor mesons within the QCD sum rules method in
\cite{Rugr06}.

In the present work we calculate the magnetic dipole moments of these
negative parity tensor mesons in the framework of the light cone QCD sum
rules method (for more about the light cone QCD sum rules method, see
\cite{Rugr07} and \cite{Rugr08}).

The paper is organized as follows. Section 2 is devoted to the derivation of
the light cone QCD sum rules for the magnetic dipole moment of the negative
parity mesons $\bar{q}q$, $\bar{q}s$, $\bar{s}s$, $\bar{q}c$, $\bar{s}c$,
$\bar{q}b$ and $\bar{s}b$. In section 3, numerical analysis of the obtained
sum rules for the dipole moments of the $2^{--}$ tensor mesons is performed.
This section also contains the discussions, and brief summary of the present
study.

\section{Theoretical framework}

In this section we derive the light cone sum rules for the magnetic dipole
moments of the negative parity
tensor mesons. For this goal we consider the 3--point correlation function,
\bea
\label{eugr01}
\Pi_{\mu\nu\rho\alpha\beta} (p,q) = - \int d^4x \int d^4y e^{i(px+qy)}
\lla 0 \vel {\cal T} \Big\{ j_{\mu\nu} (0) j_\rho^{el}(y) \bar
j_{\alpha\beta}(x) \Big\}
\ver 0 \rra~,
\eea
where $j_{\mu\nu}$ is the tensor interpolating current with $J^{PC}=2^{--}$, 
\bea
\label{eugr02}  
j_{\mu\nu} = {i\over 2} \left[ \bar{q}_1(x) \gamma_\mu \gamma_5 \olra{\cal
D}_\nu q_2(x) + \bar{q}_2 (x) \gamma_\nu \gamma_5 \olra{\cal D}_\mu q_1(x)
\right]~.
\eea
The electromagnetic current $j_\rho^{el}$ in Eq. (\ref{eugr01}) is
defined as,
\bea
j_\rho^{el} = e_{q_1} \bar{q}_1 \gamma_\rho q_1 + e_{q_2} \bar{q}_2
\gamma_\rho q_2~, \nnb
\eea
where $e_{q_i}$ is the electric charge of the corresponding quark. The
momenta $p$ and $q$ are carried by the currents  $j_{\mu\nu}$ and
$j_\rho^{el}$, respectively.

The covariant derivative $\olra{\cal D}$ is defined as
\bea
\label{eugr03}
\olra{\cal D}_\mu (x) = {1\over 2} \Big[
\ora{\cal D}_\mu (x) -
\ola{\cal D}_\mu (x) \Big]~,
\eea
where
\bea
\ora{\cal D}_\mu (x) \es \ora{\partial}_\mu (x) - i
{g\over 2} \lambda^a A_\mu^a (x) ~, \nnb \\
\ola{\cal D}_\mu (x) \es \ola{\partial}_\mu (x) + i
{g\over 2} \lambda^a A_\mu^a (x)~.\nnb
\eea
Here $A_\mu^a (x)$ is the gluon
field, satisfying the Fock-Schwinger gauge condition $x^\mu A_\mu^a (x)=0$,
which we have used in the present work, and  $\lambda^a$ are the
Gell-Mann matrices.

The correlation function can be rewritten in terms of the external background
electromagnetic field. For this goal it is necessary to introduce plane wave
electromagnetic field,
\bea
F_{\mu\nu} = i (q_\mu\varepsilon_\nu-q_\nu\varepsilon_\mu)
e^{iqx}~,\nnb
\eea
where $\varepsilon_\mu$ is the polarization vector, $q_\mu$
is the four--momentum vector of the background electromagnetic field,
and the
radiated photon can be absorbed into the background field. This
allows us to rewrite the correlation function as,
\bea
\label{eugr04}
\varepsilon^\rho \, \Pi_{\mu\nu\rho\alpha\beta} = i \int d^4x e^{ipx}
\lla 0 \vel T \Big\{ j_{\mu\nu} (x) \bar{j}_{\alpha\beta} (0) \ver 0
\rra_F~,
\eea
where the subindex $F$ means that the vacuum expectation value is calculated
in the background electromagnetic field.

Note that the correlation function given in Eq. (\ref{eugr01}) can be obtained
from Eq. (\ref{eugr04}) by expanding it in powers of $F_{\mu\nu}$, and
taking only the terms linear in $F_{\mu\nu}$ (more technical details about the
external background field method can be found in \cite{Rugr09,Rugr10}). The
main advantage of using the background field method is that it separates the
hard and soft contributions in a gauge invariant way. Hence, the main object in
our discussion is the correlation function given in Eq. (\ref{eugr04}). Here
a cautionary note is in order. Since the current $J_{\mu\nu}$ contains
derivatives, we first replace $\bar{J}_{\alpha\beta}(0)$ in Eq. (\ref{eugr04})
with $\bar{J}_{\alpha\beta}(y)$, and after carrying out the calculations we
set the variable $y$ to zero.

In order to obtain the sum rules for the dipole magnetic moment the tensor
mesons, one should insert the spin--2 mesons into the correlation function,
as the result of which we obtain,
\bea
\label{eugr05}
\Pi_{\mu\nu\alpha\beta}  \es 
i {\lla 0 \vel j_{\mu\nu} \ver
T(p,\epsilon) \rra \over p^2-m_T^2} \lla T(p,\epsilon) \vel
\right. T(p+q,\epsilon) \rra_F
{\lla T(p+q,\epsilon) \vel \bar{j}_{\alpha\beta} \ver
0 \rra \over (p+q)^2-m_T^2}+ \cdots~.
\eea
The matrix element $\lla 0 \vel j_{\mu\nu} \ver T(p,\epsilon) \rra$
is defined as,
\bea
\label{eugr06}
\lla 0 \vel j_{\mu\nu} \ver T(p,\epsilon) \rra = f_T m_T^3
\epsilon_{\mu\nu}~,
\eea
where $f_T$ is the decay constant of the tensor meson, and
$\epsilon_{\mu\nu}$ is the polarization tensor.

In the presence of the background electromagnetic field, the transition matrix
element $\lla T(p\epsilon) \ve T(p+q,\epsilon) \rra_F$ is parametrized
as follows:
\bea   
\label{eugr07}
\lla T(p,\epsilon) \ve T(p+q,\epsilon) \rra_F \es
\epsilon_{\alpha^\prime\beta^\prime}^\ast (p) \Bigg\{ 2
(\varepsilon^\prime p)
\Bigg[ g^{\alpha^\prime\rho} g^{\beta^\prime\sigma} F_1(q^2)
- g^{\beta^\prime\sigma} {q^{\alpha^\prime} q^\rho \over 2 m_T^2} F_3(q^2) +
{q^{\alpha^\prime} q^\rho
\over 2 m_T^2}\, {q^{\beta^\prime} q^\sigma \over 2 m_T^2} F_5(q^2) \Bigg] \nnb \\
\ar (\varepsilon^{\prime\sigma} q^{\beta^\prime} - \varepsilon^{\prime\beta^\prime}
q^\sigma)
\Bigg[ g^{\alpha^\prime\rho} F_2(q^2) - {q^{\alpha^\prime} q^\rho \over 2 m_T^2}
F_4(q^2) \Bigg]
\Bigg\} \epsilon_{\rho\sigma} (p+q)~,
\eea 
where $F_i(q^2)$ are the form factors.

In analysis of the experimental data it is more convenient to use
the form factors of a definite multipole in a given reference frame.
Relations between these two sets of form factors for the arbitrary integer,
and half--integer spin are derived in \cite{Rugr11}, and
the relations for the real photon case are:
\bea
\label{eugr08}
F_1(0) \es G_{E_0}(0)~, \nnb \\
F_2(0) \es G_{M_1}(0)~, \nnb \\
F_3(0) \es - 2 G_{E_0}(0) +  G_{E_2}(0) + G_{M_1}(0)~, \nnb \\
F_4(0) \es - G_{M_1}(0) + G_{M_3}(0)~, \nnb \\
F_5(0) \es G_{E_0}(0) - [ G_{E_2}(0) + G_{M_1}(0) ] + G_{E_4}(0) +
G_{M_3}(0)~,
\eea
where $G_{E_\ell}(0)$ and $G_{M_\ell}(0)$ are the electric and magnetic
multipoles.
Substituting these form factors in Eq. (\ref{eugr07}) we get,
\bea
\label{eugr09}
\lla T(p,\epsilon)\vel T(p+q,\epsilon^\prime) \right.\rra_F \es
\epsilon_{\alpha^\prime\beta^\prime}^\ast (p) \Bigg\{ 2 (\varepsilon^\prime \cdot p) 
\Bigg[ g^{\alpha^\prime\lambda} g^{\beta^\prime\sigma} G_{E_0} - 
{q^{\alpha^\prime} q^\lambda \over 2 m_T^2} g^{\beta^\prime\sigma} 
(-2 G_{E_0} + G_{E_2} + G_{M_1} ) \nnb \\
\ar {q^{\alpha^\prime} q^\lambda \over 2 m_T^2} {q^{\beta^\prime} q^\sigma
\over 2 m_T^2} (  G_{E_0} - G_{E_2} - G_{M_1} + G_{E_4} + G_{M_3} )
\Bigg] \nnb \\
\ar (\varepsilon^{\prime \sigma} q^{\beta^\prime} -
\varepsilon^{\prime\beta^\prime} q^\sigma ) \Bigg[ g^{\alpha^\prime\lambda} G_{M_1} -
{q^{\alpha^\prime} q^\lambda \over 2 m_T^2} ( - G_{M_1} + G_{M_3} ) \Bigg]
\Bigg\} \epsilon^{\lambda\sigma} (p+q)~.
\eea

In order to obtain the correlation function from the physical side, we 
substitute Eqs. (\ref{eugr06}) and (\ref{eugr09}) into Eq. (\ref{eugr05}), and
perform summation over the spins of the tensor particles by using,
\bea
\label{eugr10}
\epsilon_{\mu\nu} (p) \epsilon_{\alpha\beta}^\ast (p) \es
{1\over 2} {\cal P}_{\mu\alpha} {\cal P}_{\nu\beta}+
{1\over 2} {\cal P}_{\mu\beta} {\cal P}_{\nu\alpha}-
{1\over 3} {\cal P}_{\mu\nu} {\cal P}_{\alpha\beta}~,
\eea
where
\bea
{\cal P}_{\mu\nu} = -g_{\mu\nu} + {p_\mu p_\nu \over m_T^2},~\nnb
\eea
we obtain,
\bea
\label{eugr11}
&&\Pi_{\mu\nu\rho\alpha\beta} (p,q) \, \varepsilon^\rho =
{m_{T_Q}^6 g_{T_Q}^2 \over (p^2-m_{T_Q}^2) [(p+q)^2-m_{T_Q}^2]}
\Big\{
{1\over 2} {\cal P}_{\mu\alpha^\prime}(p) {\cal P}_{\nu\beta^\prime}(p) +
{1\over 2} {\cal P}_{\mu\beta^\prime}(p) {\cal P}_{\nu\alpha^\prime}(p) \nnb \\
\ek {1\over 3} {\cal P}_{\mu\nu}(p) {\cal P}_{\alpha^\prime\beta^\prime}(p)
\Big\} \times
\Bigg\{2 (p\mcdot \varepsilon) \Bigg[ g^{\alpha^\prime\lambda}
g^{\beta^\prime\sigma} G_{E_0}(0)
- g^{\beta^\prime\sigma} {q^{\alpha^\prime} q^\lambda \over 2
m_{T_Q}^2} \Big( - 2 G_{E_0}(0)+G_{E_2}(0) \nnb \\
\ar G_{M_1}(0)\Big)
+ {q^{\alpha^\prime} q^\lambda \over 2 m_{T_Q}^2}\, {q^{\beta^\prime}
q^\sigma \over
2 m_{T_Q}^2} \Big(G_{E_0}(0) - [G_{E_2}(0) + G_{M_2}(0)] +
G_{E_4}(0) + G_{M_3}(0)\Big)
\Bigg]\nnb \\
\ar \Big(\varepsilon^\sigma q^{\beta^\prime} -
\varepsilon^{\beta^\prime} q^\sigma \Big)
\Bigg[ g^{\alpha^\prime\lambda} G_{M_1}(0) - {q^{\alpha^\prime} q^\lambda \over 2 m_{T_Q}^2}
\Big( - G_{M_1}(0) + G_{M_2}(0)\Big) \Bigg]
\Bigg\} \nnb \\
\cp \Big\{
{1\over 2} {\cal P}_{\lambda\alpha}(p+q) {\cal P}_{\sigma\beta}(p+q) +   
{1\over 2} {\cal P}_{\lambda\beta}(p+q) {\cal P}_{\alpha\sigma}(p+q) -
{1\over 3} {\cal P}_{\lambda\sigma}(p) {\cal P}_{\alpha\beta}(p+q)
\Big\}~.  
\eea
One can easily see that the expression of the correlation function contains   
many independent structures, and any one of these structures 
can be used in the analysis of the multipole moments of the tensor mesons.
In this work we restrict ourselves to calculate the magnetic 
dipole form factor only
and for this aim we choose the structure $(\varepsilon^{\prime\beta} q^\nu -
\varepsilon^{\prime\nu} q^\beta) g^{\mu\alpha}$, whose coefficient is,
\bea
\label{eugr12}
\Pi = {m_T^6 g_T^2 \over (p^2-m_T^2) [(p+q)^2 - m_T^2]}
\Bigg\{ {1\over 4} G_{M_1} + 
\mbox{\rm other structures} \Bigg\} + \cdots ~.
\eea
The choice of of this structure is dictated by the fact that it
does not contain any contribution from the contact terms (see \cite{Rugr12}).

Using the operator product expansion (OPE), we calculate the correlation
from the QCD side in deep Euclidean region where $p^2 \rar -\infty$ and
$(p+q)^2 \rar -\infty$. After contracting all quark fields we obtain,
\bea
\label{eugr13}
\Pi_{\mu\nu\alpha\beta} \es
{-i \over 16} \int e^{ip\cdot x} e^{-i(p+q)\cdot y} d^4x\,d^4y
\lla 0 \vel \Big\{
S_{q_1} (y-x) \gamma_\mu \gamma_5 \Big[
 \ora{\partial}_\nu (x) \ora{\partial}_\beta (y) -
 \ora{\partial}_\nu (x) \ola{\partial}_\beta (y) \right. \right.\nnb \\
\ek \left. \left. \ola{\partial}_\nu (x) \ora{\partial}_\beta (y)  +
 \ola{\partial}_\nu (x) \ola{\partial}_\beta (y) \Big]
S_{q_2} (x-y) \gamma_\alpha \gamma_5 \Big\}\ver 0 \rra_F +
\{ \beta \leftrightarrow \alpha \} + \{ \nu \leftrightarrow \mu \} \nnb \\
\ar \{ \beta \leftrightarrow \alpha,~ \nu \leftrightarrow \mu \}~.
\eea
As has been noted, we set $y=0$ after performing the derivatives
with respect to $y$.
It follows from Eq. (\ref{eugr13}) that in calculation of the correlation
function the quark operators are needed. The expression of the light quark
operator is given as,
\bea
\label{eugr14}
S_q(x-y) \es S^{free} (x-y) - {\langle qq\rangle \over 12} \Bigg[1 -i {m_q
\over
4} (\not\!{x}-\not\!{y}) \Bigg] - {(x-y)^2 \over 192} m_0^2 \langle
qq\rangle
\Bigg[1 -i {m_q \over 6} (\not\!{x}-\not\!{y}) \Bigg] \nnb \\
\ek i g_s \int_0^1 du \Bigg\{
{\not\!x - \not\!{y} \over 16 \pi^2 (x-y)^2} G_{\mu\nu}(u(x-y))
\sigma^{\mu\nu}
- u (x^\mu-y^\mu) G_{\mu\nu}(u(x-y)) \gamma^\nu \nnb \\
\cp {i \over 4 \pi^2 (x-y)^2}
- i {m_q \over 32 \pi^2} G_{\mu\nu}(u(x-y)) \sigma^{\mu\nu} \Bigg[
\ln\left( -{(x-y)^2
\Lambda^2 \over u} + 2 \gamma_E \right) \Bigg] \Bigg\}~,
\eea
where
\bea
S_q^{free} (x-y) \es {i (\not\!x-\not\!y) \over 2 \pi^2 (x-y)^4} -
{m_q \over 4 \pi^2 (x-y)^2}~,\nnb
\eea
is the free quark operator,
and $\Lambda=(0.5-1.0)~GeV$ \cite{Rugr14} is the scale parameter
separating the perturbative and nonperturbative regions.
It should be remembered that the light cone expansion of the light quark
propagator is obtained in \cite{Rugr13}, which gets contributions from
nonlocal three $\bar{q}G q$, and four-particle $\bar{q}q\bar{q}q$,
$\bar{q}G^2 q$ operators, where $G_{\mu\nu}$ is the gluon strength tensor.
Expansion in in conformal spin proves that the contributions coming from
four-particle operators are small and can be neglected \cite{Rugr15}.

The expression of the
complete heavy quark propagator in the coordinate space is given as,
\bea
\label{eugr15}
S_Q(x) \es S_Q^{free} -
{g_s \over 16 \pi^2} \int_0^1 du
G_{\mu\nu}(u(x-y)) \Bigg( i \Big[\sigma^{\mu\nu} (\not\!{x}-\not\!{y}) +
(\not\!{x}-\not\!{y})
\sigma^{\mu\nu}\Big] \nnb \\
\cp{K_1 (m_Q\sqrt{-(x-y)^2})\over \sqrt{-(x-y)^2}}
+ 2 \sigma^{\mu\nu} K_0(m_Q\sqrt{-(x-y)^2})\Bigg)\Bigg\} +\cdots~,
\eea
respectively, where $K_i(m_Q\sqrt{-x^2})$ are
the modified Bessel functions. The
free part of the heavy quark propagator has the following form:
\bea
S_Q^{free} = {m_Q^2 \over 4 \pi^2} \Bigg\{ {K_1(m_Q\sqrt{-(x-y)^2}) \over
\sqrt{-(x-y)^2}} + i {(\not\!{x}-\not\!{y}) \over -(x-y)^2}
K_2(m_Q\sqrt{-(x-y)^2}) \Bigg\}~.\nnb
\eea
The correlation function contains short distance
(perturbative), and long distance (nonperturbative) contributions. The
nonperturbative contribution can be obtained
by making the replacement, 
\bea
\label{eugr16}
S_{\mu\nu}^{ab}(x-y) \to -{1\over 4} \bar{q}^a (x) \Gamma_\rho q^b (y)
\left(\Gamma_\rho \right)_{\mu\nu},
\eea
in the light quark operator given in Eq. (\ref{eugr13}), where
$\Gamma_\rho=\left\{ 1,\gamma_5,\gamma_\mu,i\gamma_5\gamma_\mu,
\sigma_{\mu\nu}/\sqrt{2}\right\}$. Moreover, matrix elements of the nonlocal
operators, such as $\bar{q}(x) \Gamma q(y)$, $\bar{q}(x) F_{\mu\nu}\Gamma q(y)$,
and $\bar{q}(x) G_{\mu\nu} \Gamma q(y)$ appear between vacuum and photon states,
when a photon interacts with the light quark fields at large distance.
Parametrization of these matrix elements in terms of the photon distribution
amplitudes (DAs) is obtained in \cite{Rugr10},
\bea
\label{eugr17}
&&\langle \gamma(q) \vert  \bar q(x) \sigma_{\mu \nu} q(0) \vert  0
\rangle  = -i e_q \qq (\varepsilon_\mu q_\nu - \varepsilon_\nu
q_\mu) \int_0^1 du e^{i \bar u qx} \left(\chi \varphi_\gamma(u) +
\frac{x^2}{16} \mathbb{A}  (u) \right) \nnb \\ &&
-\frac{i}{2(qx)}  e_q \qq \left[x_\nu \left(\varepsilon_\mu - q_\mu
\frac{\varepsilon x}{qx}\right) - x_\mu \left(\varepsilon_\nu -
q_\nu \frac{\varepsilon x}{q x}\right) \right] \int_0^1 du e^{i \bar
u q x} h_\gamma(u)
\nnb \\
&&\langle \gamma(q) \vert  \bar q(x) \gamma_\mu q(0) \vert 0 \rangle
= e_q f_{3 \gamma} \left(\varepsilon_\mu - q_\mu \frac{\varepsilon
x}{q x} \right) \int_0^1 du e^{i \bar u q x} \psi^v(u)
\nnb \\
&&\langle \gamma(q) \vert \bar q(x) \gamma_\mu \gamma_5 q(0) \vert 0
\rangle  = - \frac{1}{4} e_q f_{3 \gamma} \epsilon_{\mu \nu \alpha
\beta } \varepsilon^\nu q^\alpha x^\beta \int_0^1 du e^{i \bar u q
x} \psi^a(u)
\nnb \\
&&\langle \gamma(q) | \bar q(x) g_s G_{\mu \nu} (v x) q(0) \vert 0
\rangle = -i e_q \qq \left(\varepsilon_\mu q_\nu - \varepsilon_\nu
q_\mu \right) \int {\cal D}\alpha_i e^{i (\alpha_{\bar q} + v
\alpha_g) q x} {\cal S}(\alpha_i)
\nnb \\
&&\langle \gamma(q) | \bar q(x) g_s \tilde G_{\mu \nu} i \gamma_5 (v
x) q(0) \vert 0 \rangle = -i e_q \qq \left(\varepsilon_\mu q_\nu -
\varepsilon_\nu q_\mu \right) \int {\cal D}\alpha_i e^{i
(\alpha_{\bar q} + v \alpha_g) q x} \tilde {\cal S}(\alpha_i)
\nnb \\
&&\langle \gamma(q) \vert \bar q(x) g_s \tilde G_{\mu \nu}(v x)
\gamma_\alpha \gamma_5 q(0) \vert 0 \rangle = e_q f_{3 \gamma}
q_\alpha (\varepsilon_\mu q_\nu - \varepsilon_\nu q_\mu) \int {\cal
D}\alpha_i e^{i (\alpha_{\bar q} + v \alpha_g) q x} {\cal
A}(\alpha_i)
\nnb \\
&&\langle \gamma(q) \vert \bar q(x) g_s G_{\mu \nu}(v x) i
\gamma_\alpha q(0) \vert 0 \rangle = e_q f_{3 \gamma} q_\alpha
(\varepsilon_\mu q_\nu - \varepsilon_\nu q_\mu) \int {\cal
D}\alpha_i e^{i (\alpha_{\bar q} + v \alpha_g) q x} {\cal
V}(\alpha_i) \nnb \\ && \langle \gamma(q) \vert \bar q(x)
\sigma_{\alpha \beta} g_s G_{\mu \nu}(v x) q(0) \vert 0 \rangle  =
e_q \qq \left\{
        \left[\left(\varepsilon_\mu - q_\mu \frac{\varepsilon x}{q x}\right)\left(g_{\alpha \nu} -
        \frac{1}{qx} (q_\alpha x_\nu + q_\nu x_\alpha)\right) \right. \right. q_\beta
\nnb \\ && -
         \left(\varepsilon_\mu - q_\mu \frac{\varepsilon x}{q x}\right)\left(g_{\beta \nu} -
        \frac{1}{qx} (q_\beta x_\nu + q_\nu x_\beta)\right) q_\alpha
\nnb \\ && -
         \left(\varepsilon_\nu - q_\nu \frac{\varepsilon x}{q x}\right)\left(g_{\alpha \mu} -
        \frac{1}{qx} (q_\alpha x_\mu + q_\mu x_\alpha)\right) q_\beta
\nnb \\ &&+
         \left. \left(\varepsilon_\nu - q_\nu \frac{\varepsilon x}{q.x}\right)\left( g_{\beta \mu} -
        \frac{1}{qx} (q_\beta x_\mu + q_\mu x_\beta)\right) q_\alpha \right]
   \int {\cal D}\alpha_i e^{i (\alpha_{\bar q} + v \alpha_g) qx} {\cal T}_1(\alpha_i)
\nnb \\ &&+
        \left[\left(\varepsilon_\alpha - q_\alpha \frac{\varepsilon x}{qx}\right)
        \left(g_{\mu \beta} - \frac{1}{qx}(q_\mu x_\beta + q_\beta x_\mu)\right) \right. q_\nu
\nnb \\ &&-
         \left(\varepsilon_\alpha - q_\alpha \frac{\varepsilon x}{qx}\right)
        \left(g_{\nu \beta} - \frac{1}{qx}(q_\nu x_\beta + q_\beta x_\nu)\right)  q_\mu
\nnb \\ && -
         \left(\varepsilon_\beta - q_\beta \frac{\varepsilon x}{qx}\right)
        \left(g_{\mu \alpha} - \frac{1}{qx}(q_\mu x_\alpha + q_\alpha x_\mu)\right) q_\nu
\nnb \\ &&+
         \left. \left(\varepsilon_\beta - q_\beta \frac{\varepsilon x}{qx}\right)
        \left(g_{\nu \alpha} - \frac{1}{qx}(q_\nu x_\alpha + q_\alpha x_\nu) \right) q_\mu
        \right]
    \int {\cal D} \alpha_i e^{i (\alpha_{\bar q} + v \alpha_g) qx} {\cal T}_2(\alpha_i)
\nnb \\ &&+
        \frac{1}{qx} (q_\mu x_\nu - q_\nu x_\mu)
        (\varepsilon_\alpha q_\beta - \varepsilon_\beta q_\alpha)
    \int {\cal D} \alpha_i e^{i (\alpha_{\bar q} + v \alpha_g) qx} {\cal T}_3(\alpha_i)
\nnb \\ &&+
        \left. \frac{1}{qx} (q_\alpha x_\beta - q_\beta x_\alpha)
        (\varepsilon_\mu q_\nu - \varepsilon_\nu q_\mu)
    \int {\cal D} \alpha_i e^{i (\alpha_{\bar q} + v \alpha_g) qx} {\cal T}_4(\alpha_i)
                        \right\} \nnb \\
&&\langle \gamma(q) \vert \bar q(x) e_q F_{\mu\nu} (vx) q(0) \vert 0 \rangle =
-i e_q \qq \left(\varepsilon_\mu q_\nu - \varepsilon_\nu q_\mu \right)
\int {\cal D}\alpha_i e^{i (\alpha_{\bar q} + v \alpha_g) q x}
{\cal S}^\gamma (\alpha_i)\nnb \\
&&\langle \gamma(q) \vert \bar q(x) \sigma_{\alpha\beta} F_{\mu\nu}
(vx) q(0) \vert 0 \rangle = e_q \qq {1\over qx} (q_\alpha x_\beta - q_\beta x_\alpha)
\left(\varepsilon_\mu q_\nu - \varepsilon_\nu q_\mu \right) \nnb \\
&&\times \int {\cal D}\alpha_i e^{i (\alpha_{\bar q} + v \alpha_g) q x}
{\cal T}_4^\gamma (\alpha_i)\nnb \,,
\eea
where $\varphi_\gamma(u)$ is the leading twist--2, $\psi^v(u)$,
$\psi^a(u)$, ${\cal A}$ and ${\cal V}$ are the twist--3, and
$h_\gamma(u)$, $\mathbb{A}$, ${\cal S}$, $\widetilde{\cal S}$,
${\cal S}^\gamma$,
${\cal T}_i$ ($i=1,~2,~3,~4$), ${\cal T}_4^\gamma$ are the
twist--4 photon DAs, $\chi$ is the magnetic susceptibility, and
the measure ${\cal D} \alpha_i$ is defined as
\bea
\int {\cal D} \alpha_i = \int_0^1 d \alpha_{\bar q} \int_0^1 d
\alpha_q \int_0^1 d \alpha_g \delta(1-\alpha_{\bar
q}-\alpha_q-\alpha_g)~.\nnb
\eea
Equating the coefficients of the Lorentz structure
$(\varepsilon^\beta q^\nu - \varepsilon^\nu q^\beta) g_{\mu\alpha}$ from
both representations of the correlation function,
the sum rules for the magnetic moments of the negative parity tensor mesons
are obtained. 
To suppress the contributions of the higher states and
continuum, double Borel transformation with respect to the
variables $-p^2$ and $-(p+q)^2$ is performed. After this transformation,
finally, the magnetic moment of the negative parity tensor mesons is
obtained whose explicit expressions are given as, 

\begin{itemize}
\item {\bf Light tensor mesons} 
\end{itemize}
\bea
\label{eugr18}
&&{m_T^6 g_T^2 \over 4} e^{-m_{T}^2/M^2} G_{M_1}(q^2=0) = \nnb \\
&&{1\over 24} M^2 E_0(x) \Big[ e_{q_1} m_{q_2} \langle \bar{q}_1 q_1 \rangle
\Big( \Bbb{A}(u_0) + 4 u_0
j_1(h_\gamma)  - 2 \widetilde{j}_1(h_\gamma) \Big) \nnb \\
\ek e_{q_2} m_{q_1} \langle \bar{q}_2 q_2 \rangle \Big( \Bbb{A}(\bar{u}_0) +
2 j_2(h_\gamma)  + \widetilde{j}_2(h_\gamma) \Big) \Big] \nnb \\
\ek {1\over 48 \pi^2} M^4 E_1(x) (3-4 u_0) (e_{q_1} - e_{q_2}) m_{q_1} m_{q_2} \nnb \\
\ek {1\over 48} f_{3\gamma} M^4 E_1(x) \Big[e_{q_2} \Big( 8 j_2(\psi_v) -
\psi_a(\bar{u}_0) + 4 \psi_v(\bar{u}_0)
+\psi_a^\prime(\bar{u}_0) \Big) \nnb \\
\ek e_{q_1} \Big(8 j_1(\psi_v) -
\psi_a(u_0) + 2 u_0 (4 \psi_v(u_0)
-\psi_a^\prime(u_0))\Big)\Big] \nnb \\
\ek {1\over 240 \pi^2} M^6 E_2(x) (5 - 18 u_0) (e_{q_1} - e_{q_2}) \nnb \\
\ek {1\over 72} m_0^2 [e_{q_2} \langle \bar{q}_1 q_1 \rangle (m_{q_1} -
3 m_{q_2}) + e_{q_1} \langle \bar{q}_2 q_2 \rangle (3 m_{q_1} -
m_{q_2})]~.
\eea
\begin{itemize}
\item{\bf Heavy tensor mesons}
\end{itemize}
%
%
\bea
&&{m_{T_Q}^6 g_{T_Q}^2 \over 4} e^{-m_{T_Q}^2/M^2} G_{M_1}(q^2=0) = \nnb \\
&&{1 \over 1152 \pi^2} 
\Big[e_q \GG M^2 \left(2 m_Q^2 {\cal I}_2 - m_Q^4 {\cal I}_3\right)\Big]
- {e^{-m_Q^2/M^2}\over 3456 m_Q \pi^2} 
 M^2 \Big\{9 m_Q \left(e_q \GG - 96 e_Q m_Q \pi^2 \qq\right) \nnb \\
\ek 4 e_q \pi^2 \Big[18 m_Q^2 \qq \left(\mathbb{A} (u_0) + 2
\widetilde{j}_1(h_\gamma) + 4 \widetilde{j}_2(h_\gamma)\right) +
\GG \qq \chi \varphi_\gamma (u_0) - 36 f_{3\gamma} m_Q^3 \psi^a (u_0)\Big]\Big\} \nnb \\
\ar {1\over 32 \pi^2} e_Q m_Q^4 M^4 \left({\cal I}_2 - m_Q^2 {\cal I}_3\right) +
{e^{-m_Q^2/M^2}\over 96}
e_q M^4 \Big\{8 f_{3\gamma} \widetilde{j}_1(\psi^v) - 8 m_Q \qq
\chi \varphi_\gamma (u_0) \nnb \\
\ek f_{3\gamma} \Big[6 \psi^a (u_0) - 4 \psi^v(u_0) +
\psi^{a\prime}(u_0)\Big]\Big\} +
{e^{-m_Q^2/M^2}\over 16 \pi^2} e_q M^6 \nnb \\
\ar {1\over 32\pi^2}
m_Q^2 M^6 \Big[2 e_Q {\cal I}_2 - 3 e_Q m_Q^2 {\cal I}_3 - 4 e_Q m_Q^4 {\cal I}_4 - 
2 e_q m_Q^4 {\cal I}_4 - 2 (e_Q - e_q) m_Q^6 {\cal I}_5 \Big] \nnb \\
\ek {e^{-m_Q^2/M^2}\over 6912 M^2}
m_Q \Big\{432 e_Q m_0^2 m_Q^2 \qq - e_q \GG \Big[4 \qq \mathbb{A} (u_0) - 
     4 (5 - 4 u_0) \qq \widetilde{j}_1(h_\gamma) \nnb \\
\ek 40 \qq \widetilde{j}_2(h_\gamma) + 
     m_Q \Big(8 m_Q \qq \chi \varphi_\gamma (u_0) - f_{3\gamma} (8 \widetilde{j}_1(\psi^v) - 
         6 \psi^a (u_0) - 4 \psi^v(u_0) - \psi^{a\prime}(u_0))\Big)\Big]\Big\} \nnb \\
\ek {e^{-m_Q^2/M^2}\over 3456 M^4}
e_q \GG m_Q^3 \Big[\qq \Big(\mathbb{A} (u_0) + 2 \widetilde{j}_1(h_\gamma) + 4
\widetilde{j}_2(h_\gamma)\Big) - 
   2 f_{3\gamma} m_Q \psi^a (u_0) \Big] \nnb \\
\ek {e^{-m_Q^2/M^2}\over 3456 M^6}
e_q \GG m_Q^5 \qq \mathbb{A} (u_0)
- {e^{-m_Q^2/M^2}\over 3456 m_Q \pi^2}
e_q \Big\{4 (\GG - 18 m_Q^4) \pi^2 \qq \mathbb{A} (u_0) \nnb \\
\ar \GG \Big[3 m_Q^3 + \pi^2 \Big(4 (2 + u_0) \qq \widetilde{j}_1(h_\gamma) +
16 \qq \widetilde{j}_2(h_\gamma) - m_Q \{12 m_Q \qq \chi \varphi_\gamma (u_0) \nnb \\ 
\ek f_{3\gamma} [12 \widetilde{j}_1(\psi^v) + 2 \psi^a (u_0) + 
           (2 - u_0) (4 \psi^v(u_0) - \psi^{a\prime}(u_0))]\}\Big)\Big]
\Big\}~,
\eea
where
\bea
u_0={M_1^2 \over M_1^2 +M_2^2}~,~~~~~M^2={M_1^2 M_2^2 \over M_1^2 +M_2^2}~.\nnb
\eea

The functions $j_n(f(u))$, and $\widetilde{j}_1(f(u))~(n=1,2)$
are defined as:
\bea
\label{nolabel}
j_1(f(u^\prime)) \es \int_{u_0}^1 du^\prime f(u^\prime)~, \nnb \\
\widetilde{j}_1(f(u^\prime)) \es \int_{u_0}^1
du^\prime (u^\prime - u_0) f(u^\prime)~, \nnb \\
j_2(f(u^\prime)) \es \int_{0}^{\bar{u}_0} du^\prime
f(u^\prime)~, \nnb \\
\widetilde{j}_2(f(u^\prime)) \es \int_{0}^{\bar{u}_0} du^\prime
(u^\prime - \bar{u}_0) f(u^\prime)~,\nnb \\
E_n(x) \es 1 - e^{-x} \sum_{k=0}^{n} {x^k\over k!} 
= {1\over n!} \int_0^x dx^\prime x^{\prime n} e^{-x^\prime}~,\nnb \\
{\cal I}_n \es \int_{m_Q^2}^{s_0} ds\, {e^{-s/M^2} \over s^n}~,\nnb
\eea
with $x=s_0/M^2$, $s_0$ being the continuum threshold, and the Borel 
parameter $M^2$ is defined as,
\bea
\label{nolabel}
M^2 = {M_1^2 M_2^2 \over M_1^2 + M_2^2}~\mbox{\rm and},~u_0={M_1^2\over
M_1^2 + M_2^2}~. \nnb
\eea


Since we
have the same heavy tensor mesons in the initial and final states, we can
set $M_1^2=M_2^2=2 M^2$, as the result of which we have,
\bea
u_0={M_1^2 \over (M_1^2+M_2^2)}={1\over 2}~.\nnb
\eea
\section{Numerical analysis}

In this section we perform the numerical analysis of the sum rules for the
magnetic dipole moments of the negative parity tensor mesons derived in the previous
section. The input parameters used in the numerical analysis are,
$\uu(\mu=1~GeV) = \dd(\mu=1~GeV)  = -(0.243)^3~GeV^3$,
$\sp \ve_{\mu=1~GeV} = (0.8 \pm 0.2) \uu(\mu=1~GeV)$,
$m_0^2=(0.8\pm 0.2)~GeV^2$ which are obtained from the mass sum rule
analysis for the light baryons \cite{Rugr16,Rugr17}, and $B$
meson \cite{Rugr18}. Furthermore, we have used the $\overline{MS}$ values of
the heavy quarks masses whose values are
$\bar{m}_b(\bar{m}_b)=(4.16\pm 0.03)~GeV$ and $\bar{m}_c(\bar{m}_c)=(1.28\pm
0.03)~GeV$ \cite{Rugr19,Rugr20}. The magnetic susceptibility of quarks is
calculated in framework of the QCD sum rules in \cite{Rugr21,Rugr22,Rugr23}.
As we have already noted, the masses of the negative parity tensor mesons
are calculated in \cite{Rugr06}, which we have used in the present work.
We further have calculated the decay constants of $J=2^-$ mesons which
are needed in the numerical analysis.

The key input parameters in the present numerical analysis are the DAs.
Below we present only the expressions of the DAs that enter to the sum rules for
the magnetic dipole moments.
\bea
\label{eugr19}
\varphi_\gamma(u) \es 6 u \bar u \Big[ 1 + \varphi_2(\mu)
C_2^{\frac{3}{2}}(u - \bar u) \Big]~,
\nnb \\
\psi^v(u) \es 3 [3 (2 u - 1)^2 -1 ]+\frac{3}{64} (15
w^V_\gamma - 5 w^A_\gamma)
                        [3 - 30 (2 u - 1)^2 + 35 (2 u -1)^4]~,
\nnb \\
\psi^a(u) \es [1- (2 u -1)^2] [ 5 (2 u -1)^2 -1 ]
\frac{5}{2}
    \Bigg(1 + \frac{9}{16} w^V_\gamma - \frac{3}{16} w^A_\gamma
    \Bigg)~,
\nnb \\
{\cal A}(\alpha_i) \es 360 \alpha_q \alpha_{\bar q} \alpha_g^2
        \Bigg[ 1 + w^A_\gamma \frac{1}{2} (7 \alpha_g - 3)\Bigg]~,
\nnb \\
{\cal V}(\alpha_i) \es 540 w^V_\gamma (\alpha_q - \alpha_{\bar q})
\alpha_q \alpha_{\bar q}
                \alpha_g^2~,
\nnb \\
h_\gamma(u) \es - 10 (1 + 2 \kappa^+ ) C_2^{\frac{1}{2}}(u
- \bar u)~,
\nnb \\
\mathbb{A}(u) \es 40 u^2 \bar u^2 (3 \kappa - \kappa^+ +1 ) +
        8 (\zeta_2^+ - 3 \zeta_2) [u \bar u (2 + 13 u \bar u) + 
                2 u^3 (10 -15 u + 6 u^2) \ln(u) \nnb \\ 
\ar 2 \bar u^3 (10 - 15 \bar u + 6 \bar u^2)
        \ln(\bar u) ]~.
\eea
The values of the constant parameters in the DAs are given as:
$\varphi_2(1~GeV) = 0$, 
$w^V_\gamma = 3.8 \pm 1.8$, $w^A_\gamma = -2.1 \pm 1.0$, 
$\kappa = 0.2$, $\kappa^+ = 0$, $\zeta_1 = 0.4$, $\zeta_2 = 0.3$, 
$\zeta_1^+ = 0$ and $\zeta_2^+ = 0$ \cite{Rugr10}.

The sum rules for the magnetic dipole moments of the
negative parity $J^{PC}=2^{--}$ tensor mesons which
are obtained in the previous section contain two more arbitrary parameters,
in addition to the input parameters summarized above: Borel mass parameter
$M^2$ and the continuum threshold $s_0$. In the analysis of the sum rules,
the working regions of these two parameters should be determined, such that
the magnetic dipole moments exhibit weak dependence on these parameters.
The working regions should
satisfy the following requirements: The upper limit of $M^2$ is determined
from the condition that the higher states contributions constitute
maximum 40\% of the perturbative ones. The lower bound of $M^2$ is obtained
by requiring that the OPE should be convergent, i.e., the higher twist
contributions should be less than the leading twist contributions. From
these conditions we have obtained the working regions of the
$J^{PC}=2^{--}$ tensor mesons, which is listed in Table 1.

\begin{table}[h]

\renewcommand{\arraystretch}{1.3}
\addtolength{\arraycolsep}{-0.5pt}
\small
$$
\begin{array}{|c|r@{\div}l|c|}
\hline \hline  
   &
\multicolumn{2}{c|}{M^2~(GeV^2)} & s_0~(GeV^2)  \\ \hline
\bar{q}q              &~\, 1.3  & 1.8  & 2.1^2  \\
\bar{q}s              &    1.4  & 2.0  & 2.2^2  \\
\bar{s}s              &    1.5  & 2.2  & 2.4^2  \\
\bar{q}c              &    2.0  & 4.0  & 3.3^2  \\
\bar{s}c              &    2.2  & 4.2  & 3.6^2  \\
\bar{q}b              &    4.5  & 7.0  & 6.2^2  \\
\bar{s}b              &    4.7  & 8.0  & 7.0^2  \\
\hline \hline

\end{array}   
$$
\caption{The working regions of the Borel parameter $M^2$, and the corresponding
values of the continuum threshold $s_0$ for the $J^{PC}=2^{--}$ tensor mesons
(these values are taken from \cite{Rugr06})}
\renewcommand{\arraystretch}{1}
\addtolength{\arraycolsep}{-1.0pt}

\end{table}
The values of the continuum threshold listed in Table 1, are determined
in \cite{Rugr06} from the
analysis of the two--point correlation function.

Using the values of the input parameters and the working regions of $M^2$
and $s_0$, the values of the magnetic dipole moments can be determined. As n
example, in Figs. (1) and (2) we present the dependence of the magnetic
dipole moments of $K_2^+$ and ${\cal D}_2^0$ mesons on $M^2$ at several fixed
values of $s_0$, respectively. It follows from these figures that the magnetic dipole
moments show weak dependence on $M^2$ in its working region. Similar
analysis for the other $J^{PC}=2^{--}$ tensor mesons are carried out whose
results are presented in Table 2.

For completeness, we also present the values of the magnetic dipole moments
for the positive parity tensor mesons in the same table. From the comparison
of the results we deduce that:

\begin{itemize}
\item In the case of light tensor mesons, the magnetic dipole moments of the
negative parity mesons are 2--5 times smaller compared to that for the
positive parity mesons.

\item For the heavy tensor mesons, however, the situation is to the contrary
namely, the magnetic moments of the negative parity tensor mesons are larger
compared to the ones for the positive parity mesons. 
\end{itemize}  

These results can be explained by the fact that, the terms proportional to
the quark mass in the expressions of the sum rules have
opposite sign. Therefore,
the contributions coming from the heavy quark mass terms are constructive
(destructive) for the negative (positive) parity tensor mesons.
Additionally, this difference can be attributed to the differences
in masses and residues of the tensor mesons of both parities.

In summary, the magnetic dipole moments of the light and heavy
$J^{PC}=2^{--}$ tensor mesons are calculated in framework of the light cone
QCD sum rules method. Comparison of the predictions for the magnetic
dipole moments of the negative and positive parity mesons is also presented.
It is observed that, the results for the magnetic dipole moments of the
negative parity light mesons are smaller compared to the ones for the corresponding
positive parity tensor mesons, while the situation is to the contrary for
the heavy tensor mesons.
   
\newpage

\begin{table}[h]

\renewcommand{\arraystretch}{1.3}
\addtolength{\arraycolsep}{-0.5pt}
\small
$$
\begin{array}{|c|c|c|}
\hline \hline  
        &{G_{M_1}(\mu_N)} & G_M(\mu_N) \\ \hline
\vspace{-0.22 cm}\mbox{Tensor} 
        &{\mbox{Negative}}   & \mbox{Positive}   \\
                 \mbox{mesons}  
        &{\mbox{parity}}     & \mbox{parity}                      \\ \hline
f_2^\pm            &~~~~  0                    & 0                \\
a_2^+              &      0.26  \pm 0.05       & 1.28 \pm 0.27    \\
a_2^0              &      0                    & 0                \\
K_2^+              &      0.26  \pm 0.05       & 0.5 \pm 0.07     \\
K_2^{0}            &    -0.015 \pm 0.003       & 0.05 \pm 0.007   \\
{\cal D}_2^0       &      1.97  \pm 0.16       & 0.3\pm 0.1       \\
{\cal D}_2^+       &      1.64  \pm 0.33       & -0.80\pm 0.08    \\
{\cal D}_{2_s}^+   &      2.5   \pm 0.6        & -0.80 \pm 0.73   \\
{\cal B}_2^+       &      1.33  \pm 0.50       & 0.62 \pm 0.11    \\    
{\cal B}_2^0       &    -1.50  \pm 0.33        & -0.20 \pm 0.05   \\    
{\cal B}_{2_s}^+   &    -0.66  \pm 0.22        & -0.23 \pm 0.05   \\ \hline \hline

\end{array}   
$$
\caption{The values of the magnetic dipole moments of the negative and
positive parity tensor mesons in units of the nuclear magneton $\mu_N$.}
\renewcommand{\arraystretch}{1}
\addtolength{\arraycolsep}{-1.0pt}

\end{table}


\newpage

\section*{Figure captions}  
{\bf Fig. (1)} Dependence of the magnetic dipole moment of the negative
parity $K_2^+$ tensor meson, on Borel
mass square $M^2$, at several fixed values of the continuum threshold,
in units of $(e/2 m_T)$.\\\\
{\bf Fig. (2)} The same as Fig. (1), but for the negative parity
${\cal D}_2^0$ tensor meson.


\newpage

\begin{figure}
\vskip 3. cm
    \includegraphics{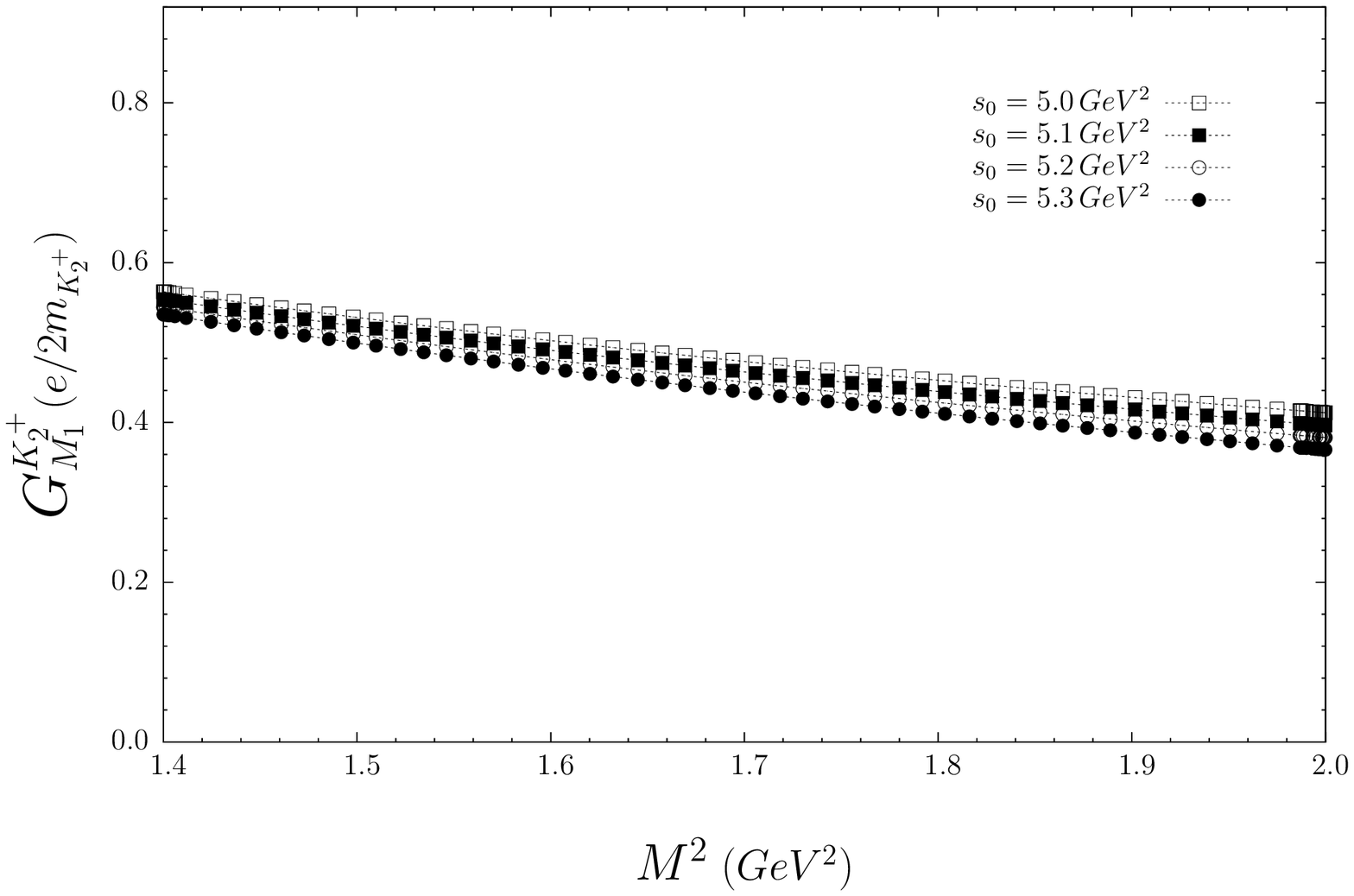}
\vskip 7.0cm
\caption{}
\end{figure}

\begin{figure}
\vskip 3. cm
    \includegraphics{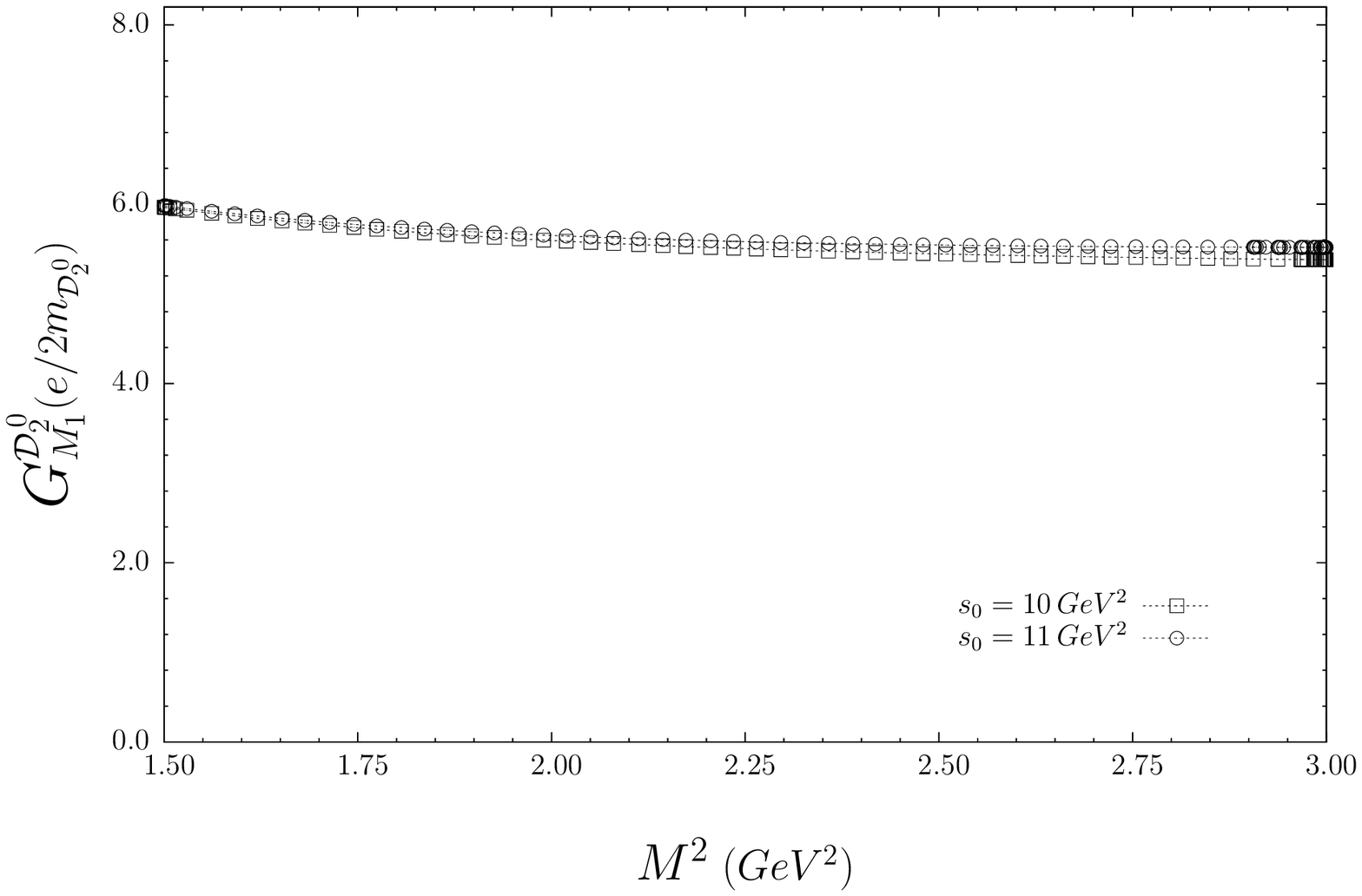}
\vskip 7.0cm
\caption{}
\end{figure}

\end {document}